\pdfoutput=1
\documentclass[runningheads]{llncs}

\usepackage[T1]{fontenc}
\usepackage[utf8]{inputenc}
\usepackage{graphicx}
\usepackage{amsmath}
\usepackage{amssymb}
\usepackage{booktabs}
\usepackage{xcolor}
\usepackage{tikz}
\usetikzlibrary{arrows.meta,positioning}
\usepackage[hidelinks]{hyperref}

\newcommand{\pend}{\textendash}

\begin{document}

\title{An Anatomy-Aware Staged Pipeline for\\Vessel-Specific Intracranial Aneurysm\\Segmentation in CT and MR Angiography}

\titlerunning{An Anatomy-Aware Staged Pipeline for Aneurysm Segmentation}

  \author{%
    Gniewosz Drwiega\inst{1}\orcidID{0000-0002-1968-2238} \and
    Wojciech Szymanski\inst{1,2}\orcidID{0009-0001-2043-9104} \and
    Marek Wodzinski\inst{1,2}\orcidID{0000-0002-8076-6246}}
  \authorrunning{G. Drwiega et al.}
  \institute{Sano -- Centre for Computational Personalised Medicine,
    International Research Foundation, Krak\'ow, Poland \and
    AGH University of Krakow, Krak\'ow, Poland\\
    \email{\{g.drwiega,w.szymanski,m.wodzinski\}@sanoscience.org}}

\maketitle

\begin{abstract}
The TopAneu 2026 challenge requires intracranial aneurysms to be segmented at the voxel level
and assigned to one of more than fifty vessel-specific classes, on CT and MR angiography alike.
We address it with a six-stage anatomy-aware pipeline rather than a single segmentation network. A
logistic-regression classifier routes each scan to a modality-specific branch; networks segment the
vasculature, first as a binary mask and then over the full 36-class arterial taxonomy; a candidate
localiser turns raw angiography and vessel priors into suspicious regions; a candidate-centred
network segments the lesion in each crop; connected components become instances; and a per-instance
classifier assigns each its anatomical location. Two principles organise the design. Whenever an
upstream model output becomes a downstream training input it is generated out-of-fold, so no stage
trains on unrealistically clean predictions of itself. And the anatomical decision is made once per
lesion rather than once per voxel, which fits the data: of 406 connected aneurysm components in the
full release, 405 carry exactly one location label. Task~1 is read off the Task~2 mask, making the
two consistent by construction. We report stage-wise performance and a five-part ablation isolating
what vessel priors and class-aware sampling contribute at each stage.
\end{abstract}

\section{Introduction}

Unruptured intracranial aneurysms occur in roughly 3\% of adults, and the decision to treat or
observe rests on a few anatomical descriptors. The cohorts underpinning current practice and the
risk scores derived from them agree that size and location dominate rupture risk and that their
effects are not separable: the same diameter carries very different risk in the posterior
circulation than in the internal carotid artery~\cite{phases}. A useful system must therefore
report not just that a lesion is present but which artery carries it.

Most methods stop short of this. Deep-learning aneurysm detection on TOF-MRA and CTA is well
established~\cite{vesselprior} and the ADAM challenge~\cite{adam} provided a reproducible
benchmark, but all treat the target as one undifferentiated lesion class.
Anatomical labelling has been pursued separately: TopCoW introduced topology-aware segmentation of
the circle of Willis~\cite{topcow} and TopBrain extended it to the whole arterial tree~\cite{topbrain}.
TopAneu 2026~\cite{topaneu} is the first benchmark to require both jointly, over more than fifty
vessel-location classes with laterality.

\begin{figure}[t]
\centering
\includegraphics[width=\textwidth]{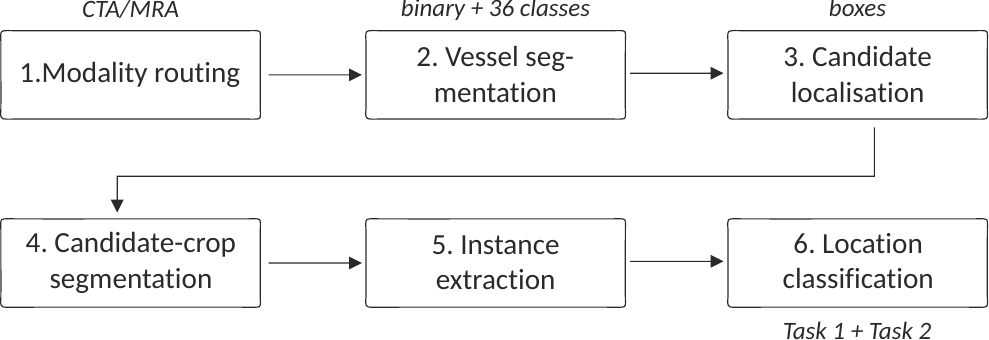}
\caption{The pipeline. Stage 4 re-segments a high-resolution crop around each candidate. Every
learned prior crossing a stage boundary is generated out-of-fold, and Task~1 is read off the Task~2
mask so the outputs cannot disagree.}
\label{fig:pipeline}
\end{figure}

Posing this as one fifty-way segmentation problem is unattractive. Aneurysm voxels are already a
vanishing fraction of an angiographic volume, and dividing that foreground across fifty long-tailed
classes leaves most locations with a handful of instances; the evidence separating them is also of
the wrong kind for a segmentation network, since what distinguishes a left from a right M1 aneurysm
is global anatomical position. We therefore decompose the task into the staged pipeline of
Fig.~\ref{fig:pipeline}, each stage operating at its own granularity and consuming only what is
available at test time. Cascading a coarse anatomical stage into a fine lesion stage is established
practice~\cite{cascadedfcn}; what is specific here is that the intermediate anatomy is itself the
label space of the final decision. Our contributions are the pipeline, a strict out-of-fold
discipline for every learned prior, and an ablation isolating what vessel anatomy and class-aware
sampling contribute.

\section{Methods}

\subsection{Data and the out-of-fold discipline}

The TopAneu release provides multi-centre CTA and MRA with aneurysm masks carrying a
vessel-location class, type masks, silver-standard vessel masks over a 36-class arterial taxonomy,
and per-case location listings~\cite{topaneu}. We used the pilot release of 98 cases (58 CTA, 40
MRA, 50 classes) and the full release of 417 --- 109 CTA, all positive, and 308 MRA of which 197
are positive --- which refines the taxonomy to 52 classes.

Vessel masks are released for training only, so every stage that consumes vessel anatomy must
consume a prediction rather than the annotation. Cases are split into five outer folds ---
stratified by aneurysm positivity for the aneurysm datasets, count-balanced for the vessel datasets
--- and each outer fold's remainder into 80/20 inner training and validation. We apply one rule
throughout: whenever an upstream model output becomes a downstream training input, it is generated
out-of-fold. Any prior consumed by a later stage therefore comes from the model whose outer fold
excluded that case, and no stage is trained on predictions of itself.

\subsection{Stage 1: modality routing}

Modality is not guaranteed to be known at inference, so it is inferred from the image. A
logistic-regression classifier over global image and header statistics --- volume shape, voxel
spacing, physical size, intensity moments and robust percentiles, with no filename information ---
routes each scan, and every subsequent stage is duplicated per modality.

\subsection{Stage 2: vessel segmentation}

Each branch segments the vasculature twice: a binary model and a multiclass model over the 36-class
taxonomy. The binary output is a cheap, robust prior; the multiclass output carries the anatomical
identity attribution needs. Whether the first helps the second is ablation A: A0 predicts the
multiclass mask from raw angiography alone, A1 from raw plus the out-of-fold binary prediction as a
second channel.

All segmentation stages are 3D full-resolution U-Nets configured per dataset by a self-configuring
framework~\cite{nnunet}, trained with deep supervision and a compound Dice and
cross-entropy objective; branches resample to their own median spacings,
$0.25\times0.43\times0.43$\,mm for CTA and $0.55\times0.30\times0.30$\,mm for MRA.

The vessel taxonomy is itself long-tailed: in the MRA split 3rd-A2 occurs in 33 of 197 training
cases and 3rd-A3 in 55, and several classes common by case presence occupy very small voxel volumes
and are learned poorly. Ablation R evaluates class-aware patch sampling. Standard nnU-Net centres
33\% of patches on foreground; the class-aware variant raises this to 75\% and, with probability
0.80, centres on one of seven small or weakly learned labels drawn uniformly, so roughly 60\% of
patches target them. Since this changes both the foreground rate and the class selection, R
separates them: default sampling (R0), the raised fraction alone (R1), and the raised fraction with
targeted selection (R2).

\subsection{Stages 3--4: candidate localisation and crop segmentation}

Rather than segmenting across the whole volume, we first localise candidates. A detector predicts a
candidate map from raw angiography and vessel priors, and connected components of the thresholded
map become bounding boxes. This raises recall and cuts the foreground/background imbalance that loss
reweighting alone does not fully remove. Ablation B varies its inputs:
raw only (B0), plus binary vessel (B1), plus multiclass (B2), plus both (B3).

A second network segments the lesion inside each crop, where the aneurysm occupies a far larger
fraction of the field of view. Ablation S varies the crop channels from raw alone (S0) through
binary vessel (S1), multiclass (S2), both (S3), and both plus the candidate map (S4); predictions
are merged back into full image space. Targets are class-collapsed to one foreground label at both
stages, so every lesion contributes to a single detection objective.

\subsection{Stages 5--6: instances, location and task outputs}

Connected components of the merged mask become aneurysm instances, which matters because a case may
carry several. The data support this granularity: the full release contains 406 components across
306 positive cases, of which 405 carry exactly one location label, so attribution is well posed as
one decision per instance.

Each instance is described by the raw crop, its instance mask, the binary and multiclass vessel
crops, and geometric features: centre of mass, volume, bounding-box size, overlap with vessel
labels, a histogram of neighbouring labels, and modality. Ablation L removes these in turn, from no
vessel information (L0) through binary features (L1) and the multiclass label histogram (L2) to the
full crop-plus-parcellation representation (L3). This matters because the label spaces diverge at
the junctions, where M1 trunk, M1 bifurcation and M1--M2 junction share a parent vessel and are
separated only by position along it.

The Task~2 mask is produced by filling each instance with its predicted location identifier, and
Task~1 is the set of unique non-zero labels in that mask.

\section{Results}

We report per-case mean and median Dice alongside \emph{pooled} Dice, summing intersections and
volumes over all cases before dividing: the latter weights every voxel equally and is therefore
dominated by large lesions. Precision and recall are pooled voxel-level. Entries marked \pend{}
will be completed for the camera-ready version.

\begin{table}[t]
\centering
\caption{Measured pilot-release results and pending ablations. All vessel priors are out-of-fold.}
\label{tab:abl}
\scriptsize
\begin{tabular}{llc}
\toprule
Stage / ablation & Configuration & Result \\
\midrule
\multicolumn{3}{l}{\emph{Measured, pilot release, out-of-fold}}\\
Vessel & binary, CTA, $n{=}58$ & 0.938 pooled Dice \\
Vessel & binary, MRA, $n{=}40$ & 0.965 pooled Dice \\
Aneurysm & full volume, MRA, $n{=}40$ & 0.378 pooled Dice, 0.262 recall \\
\midrule
\multicolumn{3}{l}{\emph{Pending, full release; variants defined in Sect.~2}}\\
A0--A1 & binary vessel prior for multiclass labelling & \pend \\
R0--R2 & foreground fraction and class-aware sampling & \pend \\
B0--B3 & vessel priors for candidate localisation & \pend \\
S0--S4 & vessel and candidate priors for crop segmentation & \pend \\
L0--L3 & vessel features for location attribution & \pend \\
E & end-to-end, CTA and MRA & \pend \\
\bottomrule
\end{tabular}
\end{table}

\textbf{Modality routing and vessel segmentation.} Under stratified five-fold cross-validation over
all 417 full-release cases the modality classifier separates CTA from MRA perfectly (balanced
accuracy and ROC AUC 1.000), so routing contributes no error downstream. Out-of-fold on the pilot
cohort the binary vessel stage reaches 0.938 pooled Dice on CTA and 0.965 on MRA, minima 0.885 and
0.913.

\textbf{Why stages 3--4 exist.} A full-volume aneurysm model on the pilot release attains 0.723
median Dice out-of-fold on MRA but only 0.378 pooled, at 0.262 recall; seven of 40 cases scored
zero. The regime is neither threshold- nor prior-limited: relaxing the threshold from 0.50 to 0.05
moves pooled recall only to 0.283, and among the worst cases all but two had 95--100\% of the lesion
inside the predicted vessel mask. What the full-volume formulation withholds is scale --- at
$128^3$ patches over tens of millions of voxels a 4\,mm$^3$ aneurysm is roughly eighty voxels.
Pooling modalities inflates the same measurement (0.782 pooled jointly against 0.282 for MRA alone)
only because CTA lesions here are one to two orders of magnitude larger, which is the empirical
case for routing by modality.

\section{Discussion and Conclusion}

The pipeline is organised around one principle: each sub-problem is solved at the granularity at
which it is well posed. Modality is an image-level property; vessels and lesions are voxel-level
structures; anatomical location belongs to a lesion rather than a voxel, and the data agree --- 405
of 406 components carry a single label, so attribution is a per-instance decision over a few hundred
examples. The second principle is the out-of-fold discipline. In a pipeline this deep, four model outputs
become training inputs downstream, and reusing folds across that boundary inflates estimates in the
way documented for non-nested model selection~\cite{varma,leakage}. One rule everywhere costs an
extra prediction pass per fold and buys estimates that do not collapse when the stages are assembled
end to end.

The ablations answer a question the pilot results raise but cannot settle: whether anatomical
information helps because it localises the search or because it identifies the vessel. B tests the
first at the detector, L the second at attribution, A whether the cheap binary prior earns its place
as input to the expensive multiclass one, and R the same of sampling. The pilot evidence already
constrains the answer --- lesions are missed even when lying almost entirely inside an accurate
vessel mask --- which is why stages 3--4 change the field of view rather than adding another prior.
In summary, we presented a six-stage anatomy-aware pipeline for the TopAneu 2026 tasks, with every
learned prior generated out-of-fold and attribution posed per lesion instance.

\subsubsection{Acknowledgements.}
This work was supported by the National Science Centre, Poland, under Grant ``MultiGeoMed''
No.\ 2024/55/D/ST6/02081. We gratefully acknowledge the Polish high-performance computing
infrastructure PLGrid, HPC Center: ACK Cyfronet AGH, for providing computational resources and
support within computational grant No.\ PLG/2025/018770.

\subsubsection{Disclosure of Interests.} The authors have no competing interests to declare.

\bibliographystyle{splncs04}
\bibliography{refs}

\end{document}